\documentclass{Rinton-P9x6}

\def\mxth{\mathsurround=0pt }
\def\xversim#1#2{\lower2.pt\vbox{\baselineskip0pt \lineskip-.5pt
  \ialign{$\mxth#1\hfil##\hfil$\crcr#2\crcr\sim\crcr}}}             
\def\simgr{\mathrel{\mathpalette\xversim >}}                                    
      
\newcommand{\beq}{\begin{equation}}                                             
\newcommand{\eeq}{\end{equation}}

\newcommand{\zpr}{\mbox{$Z'$}}
\newcommand{\mzp}{\mbox{$M_{Z'}$}}

\newcommand{\upr}{\mbox{$U(1)'$}}

\newcommand{\alsz}{\mbox{$\alpha_s(M_Z)$}}

\newcommand{\skipblk}[1]{}                                                      
\def\bqa{\begin{eqnarray}}                                                      
\def\eqa{\end{eqnarray}}                                                        
\newcommand{\sto}{\mbox{$SU(2) \x U(1)$}}                                       
                                                                                       
\newcommand{\x}{\mbox{$\times$}}

\newcommand{\lms}{\mbox{$\Lambda_{\overline{MS}}\ $}}                           
\begin{document}

\title{TeV Physics from the Top Down}

\author{Paul Langacker}

\address{Department of Physics and 
Astronomy, University of Pennsylvania \\
Philadelphia, PA 19104, USA }
%\\E-mail: pgl@electroweak.hep..upenn.edu}

\maketitle

\abstracts{Concrete semi-realistic string/M theory constructions often
predict the existence of new physics at the TeV scale, which may be different
in character from the bottom-up ideas that are motivated by specific problems
of the standard model. I describe two examples, 
heterotic and open string, of such constructions. The latter has a particularly interesting  strongly-coupled quasi-hidden
sector, which may lead to composite states at low energy and dynamical supersymmetry
breaking. General issues, such as grand unification versus direct compactification,
additional \upr \ gauge symmetries and other exotics, and flat directions are discussed. }

%\section{Motivations}

\section{Unification: from the Top Down}

Most studies of possible new physics at the TeV scale are of the bottom-up
type, motivated by attempts to solve problems or explain arbitrary features of the
standard model. However, there could also be new
physics that survives as a TeV-scale remnant of an underlying
theory, e.g., at the Planck scale, that does not directly address problems
of the standard model but which could provide an important probe of the underlying
unification or dynamics.

The  string/M theory paradigm is both ambitious and promising. It holds out a
real possibility of eventually providing a complete and consistent description
of all phenomena up to the Planck scale. Nevertheless, realization of
these goals is held up by a number of significant obstacles. These include the
need to explore the
many realms of perturbative and non-perturbative M theory, the wide variety of 
possible compactifications, and the difficulties of stabilizing the dilaton and moduli and
finding a satisfactory scheme for supersymmetry breaking.
Accounting for the vanishing or incredibly tiny
cosmological constant, $\Lambda_{\rm  cosm}$, appears to be especially
intractable.

Significant progress may have to await formal and mathematical developments in
our understanding of M theory, but it is important to carry out detailed
studies of specific constructions in parallel. 
These could suggest new TeV-scale physics,
suggest promising new directions for the exploration of M theory, allow the
development of computational techniques, and generally help us understand the
range of issues and possibilities.

It is unlikely that anyone will find a fully realistic construction in the near future,
so most studies have emphasized specific features, such as the possibility that the  
fundamental scale $M_{\rm  fund}$ is much smaller than $ M_{\rm  pl}$ (large
extra dimensions), possibilities for supersymmetry breaking and an acceptible
$\Lambda_{\rm   cosm}$, and models with dynamical 
dilaton/moduli stabilization. I will concentrate on studies 
carried out with M. Cveti\v c and others (and closely related work)~\cite{FNY}-\cite{CLW} of
constructions leading to semi-realistic four-dimensional gauge theories containing the MSSM and for which $M_{\rm  fund} \sim M_{\rm  pl}$.
 
\section{To GUT or Not To GUT}
\label{GUTsection}
 One basic question is whether (a) the underlying string theory first compactifies
 to an effective four-dimensional grand unified theory (GUT) and that
 at a lower scale the GUT is broken to the MSSM (plus possible additional gauge
 factors), or (b) whether the string theory compactifies directly to the
 MSSM (plus extensions) without the intermediate GUT stage\footnote{Grand unification
 could still exist in higher dimensions~\cite{hdgut}.}. In the latter case the traditional forms of
 gauge and Yukawa unification are typically lost, though they may still survive
 in some modified or partial form.
 
 Traditional GUTs have several major successes. These include the approximate
 success of gauge unification~\cite{gaugeunif}; the elegant explanation for the
 quantum numbers of the 15 members of each fermion family;
 a natural home for the GUT seesaw model~\cite{seesaw}, which yields the right
 general scale for neutrino masses and can incorporate the leptogenesis
 scenario for the baryon asymmetry~\cite{leptogenesis}; and a correct prediction
 for the $b$ to $\tau$ mass ratio. On the other hand, the simplest versions
 of the GUT seesaw lead to small neutrino mixings, comparable to those in the
 quark sector, contrary to observations; other fermion mass relations
 (involving the first two families) are not correct unless one complicates the
 model with higher-dimensional Higgs multiplets (and typically additional
 family symmetries); simple models may be excluded by the non-observation
 of proton decay; there are serious hierarchy problems, including the
 Higgs doublet-triplet splitting problem; one must explain the
 existence of an additional (GUT breaking) scale; and the simplest heterotic string
 models do not have the large (adjoint and other) Higgs representations
 required for the fermion spectrum and GUT breaking. For these reasons,
 one should keep an open mind as to whether a separate four-dimensional
 GUT stage is really necessary.

\section{Direct Compactification}
Direct compactifications are
constructions in which the string theory compactifies directly
to a four-dimensional theory including
the MSSM, i.e., including 
the $SU(3) \times SU(2) \times U(1)$ gauge group, 
3 families, and softly broken $N=1$ supersymmetry.

Such models often involve additional surviving gauge factors.
 These often include  quasi-hidden non-abelian groups, which may be
 candidates for dynamical supersymmetry breaking.  
 A truly hidden sector 
 would mean that no fields transform nontrivially under both the MSSM and the
hidden groups.
However, in the specific constructions
 we have examined there are a few (mixed) chiral supermultiplets which transform under
 both sectors and connect them. In addition, there are frequently additional
(non-anomalous\footnote{There are often anomalous \upr \ factors as well,
typically broken near the string scale.}) $U(1)'$ 
factors and associated \zpr \ gauge bosons, often with
family non-universal couplings which lead to flavor changing neutral 
currents (FCNC)~\cite{FCNC}.
Both the ordinary and quasi-hidden sector particles often carry \upr \ charges.
There may be additional non-abelian groups in the ordinary sector as well. 

Another common feature are new exotic chiral supermultiplets, such as
new standard model singlets (which may, however, carry \upr \
 charges), quarks or leptons with non-standard \sto  \ assignments (e.g., left-handed
 singlets and/or right-handed doublets), and
extra Higgs doublets. The particles which communicate between the gauge
sectors are typically fractionally charged (e.g., electric charge of $\pm 1/2$).  
Constructions sometimes violate $R$-parity, e.g., through
Higgs/lepton mixing.

\section{Two Examples}
In this section I describe detailed investigations of the
low energy consequences of two concrete constructions. 
Neither is fully realistic, but each predicts possible types
of new-TeV scale physics.

\subsection{A Heterotic Example}
Most of the effort in the last decade was on perturbative heterotic
constructions~\cite{FNY,CHL}. Such models involve an anomalous $U(1)_A$
symmetry, implying a constant Fayet-Iliopoulos contribution to the $U(1)_A$
$D$ term. This must be cancelled by string-scale VEVs for some
of the fields in the effective four-dimensional theory to avoid
string-scale supersymmetry breaking. A systematic procedure was
 developed in~\cite{flat} to classify all of the 
 non-abelian singlet directions that are $F$-flat and
 $D$-flat with respect to all of the gauge factors. The large VEVs break
 some of the gauge symmetries and give string-scale masses to some of the
 particles, modifying the effective four-dimensional theory from its
 initial apparent form and requiring that the interactions of
 the residual massless fields be recomputed (vacuum restabilization)~\cite{flat}.
 
 The restabilization was worked out for a number of examples in~\cite{CCEELW,clew,CFNW}.
 In particular,  a detailed analysis was made~\cite{CCEELW} of the
 flat directions involving non-abelian singlets and their consequences
 for  a construction, CHL5, due originally to 
   Chaudhuri, Hockney,  and Lykken~\cite{CHL}.
   Before restabilization, CHL5 has the
  gauge group 
  \beq \{SU(3)_C\times SU(2)_L\}_{\rm obs} 
     \times\{SU(4)_2\times SU(2)_2\}_{\rm hid}
   \times U(1)_A\times U(1)^6, \eeq 
   where the subscripts refer to the observable and quasi-hidden sector gauge groups,
 respectively. There are six non-anomalous $U(1)$ factors. 
 In addition to the MSSM fields, the particle
 content consists of  
  \bqa  & & 6 (1,2,1,1) { ({\rm extra}\ H,L)} +
  \left[(3,1,1,1) + (\bar{3},1,1,1)\right] {({\rm exotic}\ D)} \nonumber   \\   
 &+&   4 (1,2,1,2)  {\rm  (mixed)} +42 (1,1,1,1){\rm  (NA \ singlets)} \nonumber   \\
 &+& \left[ 2 (1,1,4,1) + 10 (1,1,\bar{4},1)  +
   8 (1,1,1,2) + 5 (1,1,4,2) \right.   \nonumber   \\
 &+& \left. (1,1,\bar{4},2) +
    8 (1,1,6,1) + 3 (1,1,1,3)\right]{\rm  (hidden)} 
        \eqa
%      \bqa  & & 6 (1,2,1,1) { ({\rm extra}\ H,L)} \nonumber  \\
% &+& \left[(3,1,1,1) + (\bar{3},1,1,1)\right] {({\rm exotic}\ D)} \nonumber  \\
% &+& 4 (1,2,1,2)  {\rm  (mixed)} \nonumber   \\
% &+& \left[ 2 (1,1,4,1) + 10 (1,1,\bar{4},1) \right. \nonumber  \\  
% &+& \left. 8 (1,1,1,2) + 5 (1,1,4,2) + (1,1,\bar{4},2) \right. \nonumber    \\
% &+& \left.
%    8 (1,1,6,1) + 3 (1,1,1,3)\right]{\rm  (hidden)} \nonumber   \\   
% &+& 42 (1,1,1,1){\rm  (NA \ singlets)} 
%    \eqa
The only satisfactory
 hypercharge definition is    \beq
   Y=\frac{1}{96}(-8Q_2-3Q_3-8Q_4-Q_5+Q_6), \eeq
which has Ka\v c-Moody level $11/3$, to be compared with the minimal value  $5/3$.
%value of $5/3$.

Some  features of representative flat directions after restabilization are
     \begin{itemize}
     \item There is always at least one 
     extra $Z'$ at the TeV (or possibly an intermediate) scale, with
     non family-universal couplings, leading to 
     flavor changing neutral currents. 
 
       \item  There is a quasi-hidden sector involving the $SU(4)_2\times SU(2)_2$
        group. However, a few mixed multiplets and $U(1)'$ factors connect the
        sectors. The non-abelian groups are
      {\em not} asymptotically free, so they are not a candidate for fractional charge confinement,
        supersymmetry breaking by gaugino condensation, or dilaton/moduli stabilization.     
        
      \item  There are many exotics, including
    a charge $-1/3$
    $D_{L,R}$ quark, additional Higgs/lepton doublets, non-abelian  singlets, and charge $\pm 1/2$
    particles (the mixed and some of the hidden sector states). In most 
    cases the fermions are massless or unacceptably light. There
    are extra Higgs doublets, with trilinear couplings to singlets that can generate
    non-standard off-diagonal effective $\mu$ terms, but not enough such couplings
    to be realistic.
 
       \item The higher Ka\v c -Moody embedding of hypercharge and the  exotics 
       modify the gauge unification. Using \alsz \ as input, one predicts
      $(\sin^2 \theta_W, g_2)$ $\sim (0.16, 0.48)$, which disagree with (but are not
      too far from) the experimental values (0.23, 0.65).

      \item  The primitive Yukawa couplings  at the string scale $M_s$ are not arbitrary, 
      but are related to      the gauge coupling $g$ as $g, g/\sqrt{2}$, or $0$. However, 
      after restabilization
      some couplings, originally due to higher dimensional operators,
      may be smaller, of order $g v/M_s$, where $v \sim (10^{-1}-10^{-2}) M_s$    
      is the scale of VEVs involved in the restabilization.
      The CHL5 model implies  $t-b$ universality, consistent with the data
      for large $\tan \beta$, as well as $\tau-\mu$ universality. The latter is
      unphysical and is a clear defect of the construction.  There is also a  noncanonical $b-\tau$
      relation, a semi-realistic $d$-quark texture, and no obvious mechanism for a
      $\nu$ seesaw, except possibly associated with  intermediate scale \upr \ breaking.
         
      \item  One of the flat directions involves  ${ R}_P, { L}$ and ${B}$ violation.
            
      \item When additional but ad hoc phenomenological soft supersymmetry breaking terms
       at the string scale are assumed~\cite{CCEELW}, one finds that a
     large $M_{Z'}$ of order 1 TeV is possible, the 
       sparticles  are typically heavier than in the MSSM (because the \zpr \ mass is set by
       the SUSY-breaking scale, with the smaller electroweak scale due to cancellations), and
       there is  a
       richer Higgs, neutralino, and chargino spectrum than in the MSSM~\cite{spectrum}.     
     \end{itemize}

\subsection{An Intersecting Brane (Open String) Example}
As an open string example, consider the  intersecting
$D$ brane construction of Cveti\v c, Shiu, and Uranga~\cite{CSU},
based on a $Z_2 \times Z_2$ Type IIA orientifold with three 2-tori.
It is chiral, with
an odd number of families for one tilted 2-torus. There are
 stacks of D6 branes wrapped on each $T^2$, and the model  is $N=1$ 
 supersymmetric for specific angle conditions.
 The phenomenology of the construction was considered in detail
 in~\cite{CSL1,CSL2,CLW}.
 
 The gauge group for the four dimensional effective theory is
 \beq
 SU(3)\times SU(2)\times Sp(2)_B\times Sp(2)_A \times Sp(4) \times U(1)^5, \eeq
 where the quasi hidden sector group $Sp(2)_B\times Sp(2)_A\times Sp(4) $
 is asymptotically free, allowing the possibility of charge confinement, 
 dilaton/moduli stabilization, and
dynamical supersymmetry breaking~\cite{CLW}. Two of the $U(1)$ factors are
anomalous. Their gauge bosons acquire string-scale masses, but
their symmetries restrict the Yukawa couplings. There are three
surviving non-anomalous $U(1)$'s, i.e., $Q_8 \pm Q_{8'}$
and $B-L= Q_3/3-Q_1$, where $Q_{8,8'}$ are remnants of an
underlying $Sp(8)$ symmetry, and $Q_{3,1}$ are related to an
underlying Pati-Salam type $U(4)$. 
The weak hypercharge is
\beq Y = \frac{B-L}{2} + \frac 1 2 (Q_8 + Q_{8'}). \eeq
Unfortunately, there is no simple mechanism to break the two
extra $U(1)'$ symmetries.

The chiral states associated with strings localized at intersecting
branes include three complete standard model families,
although only two of the left-handed families have Higgs Yukawa
couplings. There are actually four families of
right-handed fields (i.e., left-handed $\bar{u}, \ \bar{d}, \ \bar{e}, \ \bar{\nu}$).
These have non-universal $Q_8 - Q_{8'}$ charges, leading in general to
flavor changing neutral currents, e.g., in $B_s \rightarrow \mu^+ \mu^-$,
$B_{d} \rightarrow \phi K_{S}$,
or $\tau^+ \ra \mu^+ e^+ e^-$.

There are several exotics that are charged under both the ordinary
and quasi-hidden sectors. These include
states that transform as $(3,1,2,1,1)$ and
$(1,1,2,1,1)$ under $SU(3)\times SU(2)\times 
Sp(2)_B\times Sp(2)_A \times Sp(4)$. These are candidates to be
the exotic ($SU(2)$-singlet) left-handed partners of the extra family
of right-handed particles. Unfortunately, they have the wrong
electric charges, $Q_{EM} = Y = 1/6$ and $-1/2$, respectively,
and there is no alternative successful definition of $Y$ that
can remedy the problem. There is also an exotic $(1,2,1,1,4)$
with $Q_{EM} = \pm 1/2$. Fortunately, these fractionally-charged
states may disappear from the spectrum due to charge confinement in
the quasi-hidden sector, to be replaced with composite states 
to form the missing exotic left-handed family (see Section \ref{qhs})~\cite{CSL1}.

The particular construction actually has 24 Higgs doublet
candidates (12 $u$-type and 12 $d$-type). While additional doublets
are a real possiblility at the TeV scale, 24 is excessive, leading
to much of the difficulties with the low energy gauge couplings.
There are no candidates for chiral singlets that could generate
an effective $\mu$ term. The Yukawa interactions are discussed
in~\cite{CSL2}.

The spectrum also includes two localized singlets, two localized
$SU(2)$ triplets, and a number of unlocalized chiral singlets
and adjoints. There is no simple mechanism for giving the latter states
masses (a generic problem for such constructions.) There
are also a number of non-chiral states, which, however,
generically have string-scale masses.

The intersecting brane model does not have canonical
gauge unification, because each gauge factor is associated
with a different stack of branes\footnote{Variant constructions
yield a grand unified group and canonical gauge boundary
conditions~\cite{CSU}.}. In particular, the boundary conditions on the couplings
at the string scale depend  both on the dilaton $S$, and
also (in a non-universal way) on a modulus $\chi$,
defined as the ratio of the two radii of the first two-torus (the ratios for
the other two-tori are related by the supersymmetry condition).
The gauge couplings at low energy were
computed in~\cite{CSL1}  as functions of the modulus and dilaton, and the absolute values
were given in terms of the predicted stabilized values of $S$ and $\chi$ in~\cite{CLW}.
The predicted  standard model gauge couplings are unrealistically
small, i.e., $(\alpha_3^{-1}, \alpha^{-1}, \sin^2 \theta_W)=
(52.2,\ 525, \ 0.29),$ respectively, to be compared with the
experimental values $\sim(8.5,\ 128,\ 0.23)$. The failure is due
both to the non-canonical boundary conditions and the contributions of
the extra Higgs doublets and other exotics to the running.

\section{Things to Watch For}
Let me now describe examples of types of new physics which could emerge in specific
constructions.

\subsection{Gauge Unification?}
As described in Section \ref{GUTsection} the MSSM is consistent with the
gauge unification expected in simple four-dimensional grand unified theories
or simple perturbative heterotic string constructions.
In particular, using the precisely known values of $\alpha$ 
and the \lms \ weak angle $\hat{s}^2_Z$ (both evaluated at $M_Z$)
and assuming gauge unification, one can predict $\alsz \sim 0.130 \pm 0.010$
and the unification scale $M_G \sim 3 \times 10^{16}$ GeV~\cite{gaugeunif}.
The \alsz \ prediction agrees within 10\% with the experimental value
of $\sim 0.12$, with the discrepancy possibly due to threshold corrections or
higher dimensional operators. There is no prediction for $M_G$ in a GUT,
but one expects $M_G \sim 5 \times 10^{17}$ GeV in the simple heterotic case.
This is within 10\% when properly measured in $\ln M_G$. It is easy to find
corrections to the simple heterotic predictions, e.g., new exotic particles typically
yield $O(1)$ corrections to both \alsz \ and $\ln M_G$. The difficulty
is in understanding why the discrepancies are not larger.

More generally, string constructions with direct compactification
typically involve gauge unification, but
usually modified in form from the MSSM. For example, the boundary conditions
on the couplings may be modified due to higher Ka\v c-Moody levels
in the embeddings of the different gauge factors, or may depend
on moduli in the open-string constructions. As mentioned, exotic multiplets
may modify the runnings of the couplings. One possibility is to
consider constructions with canonical boundary conditions and no new
particles with standard model quantum numbers (there are no known
fully realistic examples). An alternative is to insist on
canonical boundary conditions and only allow exotics which form complete
$SU(5)$ multiplets (which preserves gauge unification at one-loop).
Finally, it is possible that there are cancellations between non-canonical
boundary conditions and non-standard running, in which case the apparent
success of the MSSM-type unification would be accidental\footnote{This is similar
to the non-canonical unification assumed in other theoretical
frameworks, such as large extra dimensions~\cite{LED} or 
de-construction~\cite{deconstruction}.}.

\subsection{A  TeV Scale \zpr?}
\label{zprime}
String constructions  often involve extra 
non-anomalous \upr \  gauge symmetries and associated \zpr \ gauge bosons
in the effective four-dimensional theory\footnote{Grand unified theories
also often yield extra \upr \ factors, but in that case an extra
fine tuning is required to obtain $\mzp \ll M_{ GUT}$. Theories
of dynamical symmetry breaking also often lead to an extra \upr~\cite{DSB}.}.
In both supergravity~\cite{sugra} and gauge-mediated~\cite{gaugemed} supersymmetry breaking schemes, the radiative breaking of the electroweak symmetry
often simultaneously leads to radiative
breaking of the \upr \ at the electroweak or TeV (i.e., soft
supersymmetry breaking) scale unless the breaking occurs at an
intermediate scale~\cite{intermediate} along
an $F$ and $D$ flat direction.

The \upr \ symmetry allows an elegant solution to the $\mu$ problem~\cite{muprob,Demir}.
If the superpotential $W$ contains a term $W \sim h \hat{S} \hat{H_u} \hat{H_d}$, where $\hat{S}$ is 
the superfield for a
standard model singlet that is  charged under \upr, then the \upr \ prevents an
elementary $\mu$, while  $\langle S \rangle \ne 0 $ not only breaks the \upr,
but also generates an effective $\mu_{eff} = h \langle S \rangle$. 
(This
is similar to the next to minimal supersymmetric model (NMSSM)~\cite{NMSSM}, but 
without the  domain wall problems that plague the latter~\cite{domain}. The discrete symmetry
of the NMSSM is embedded in the \upr.)

The experimental limits, both from precision and collider experiments~\cite{zprexp},
are model dependent, but
typically one requires  $\mzp > (500-800) \ GeV$ and a  $Z-Z'$ mixing   
angle     $|\delta| < {\rm  few} \x 10^{-3}$. Models with
$\mzp \simgr 10 M_Z$ and small enough mixing can be 
obtained  either by imposing a modest tuning on the parameters~\cite{Demir}, or by 
breaking the \upr \ along a nearly flat direction in a secluded
sector~\cite{ELL}. Anomaly free \upr \ models consistent with canonical
gauge unification have been constructed~\cite{JE,LW,KLLL}.

An extra TeV-scale \zpr \ is perhaps the best motivated new physics beyond
supersymmetry. The existence of a \zpr \ could have a number of other implications.
Constructions involving a low energy \upr \ typically also involve
new particles with exotic standard model quantum numbers (some of which may
be quasi-stable~\cite{EL}). The string constructions often involve \zpr \
couplings that are not family universal, leading to flavor changing
neutral current effects, especially for the third family~\cite{FCNC}.
The Higgs sector of such models is more complicated than the MSSM,
with a larger value allowed for the lightest   scalar mass. In the secluded sector case there may
be significant mixing between Higgs doublets and singlets~\cite{HLM}.
The constructions involving some tuning have a large supersymmetry breaking scale and
a non-standard sparticle spectrum~\cite{spectrum}.  The \upr \ symmetry often prevents
a traditional seesaw mechanism~\cite{seesaw} for the neutrino masses, but there are several
other possibilities for both Dirac and Majorana masses~\cite{SB,KLL}. 
There may be new sources of CP violation~\cite{KLLL,DE}.
The \upr \ models allow a strongly first order electroweak phase transition. Because of this
and the new CP phases (which may occur at tree level in the scalar sector) they allow for
 electroweak baryogenesis~\cite{KLLL} without the stringent parameter constraints
 on the MSSM~\cite{EWBMSSM}.

\subsection{Exotics}
Explicit constructions generally predict the existence of new exotic chiral supermultiplets
in the effective four-dimensional theory. These can have interesting
phenomenological consequences (and can significantly  affect gauge
unification). It is of course difficult to know to what extent these exotics should be
viewed as defects of the particular constructions, and to what extent they are 
plausible candidates for new physics. One can always give the spin-0 exotic particles
soft masses in the several hundred GeV range, but in many cases there is no
satisfactory mechanism to give large masses to the fermions.
In any case, the constructions we have studied in detail
contain too many exotics.

Many kinds of exotics are encountered.  For example, one frequently finds left-handed singlet
or right-handed doublet quarks or leptons. These can be vectorlike, i.e., 
occurring as left and right handed
pairs with the same standard model quantum numbers. In this case the electroweak
precision constraints  are weak. Familiar examples include the vectorlike $SU(2)$-singlet
charge $-1/3$ quark or the new left and right handed lepton doublets
occurring in $E_6$ grand unified theories.  Another possibility is for the
$L$-singlet and $R$-doublet particles to form partial or complete mirror families, for
which the precision constraints are strong.
Models often have many new standard model singlets, which may however be charged under \upr \
factors (see Section \ref{zprime}). There are also often additional Higgs doublets.
The Higgs doublets and singlets  may mix, leading to unusual Higgs spectra
and couplings~\cite{HLM}.
There may also be mixing between Higgs and lepton doublets, a form of $R$-parity violation.
There are usually particles that are standard model singlets that are charged under
non-abelian quasi-hidden sector gauge groups. However, the standard model and 
hidden sectors are not totally decoupled: there are typically a few particles
(often with fractional charges\footnote{Such states may 
disappear from the spectrum if the hidden sector becomes strongly coupled
(section \ref{qhs}). Otherwise, the lightest would be
stable, with cosmological implications~\cite{crypton}.}
 such as $\pm 1/2$) which couple to both sectors. The extra \upr \ factors
also may communicate between the sectors. The low energy theory is anomaly free, but
because of the various types of exotics the anomaly cancellations may be complicated.

\subsection{Flat directions}
Flat directions are an important consideration in all supersymmetric
theories. Consider, for example, a model with an additional \upr \ which
can be broken by two standard model singlets, $S_1$ and $S_2$.
Let us further assume that there are no cubic (or quadratic) terms in
the superpotential involving only the standard model singlets, i.e., the potential is 
$F$-flat at the renormalizable level\footnote{A variation is that there are cubic superpotential terms
with small coefficients, leading
to the possibility of \upr \ breaking that is naturally at the TeV scale~\cite{ELL}.}.
 The potential for
$S_{1,2}$ is then
\beq
V(S_1,S_2) = m_1^2 |S_1^2| + m_2^2 |S_2^2| + \frac{g'^2 Q'^2}{2}
(|S_1^2| - |S_2^2|)^2, \label{flatd}
\eeq
where $m_i^2$ are the soft mass-squares evaluated at the
electroweak scale, and I have assumed that $S_{1,2}$ have opposite signs for their \upr \ charges
(taken to be equal in magnitude for simplicity). $V$ is $F$ and $D$ flat along the
direction $|S_1| = |S_2| \equiv |S|$.  If $m_1^2 + m_2^2 > 0$
the flat direction will be irrelevant, and the breaking will be at or near the
electroweak scale, where additional terms such as the $F$ terms associated with
$W \sim h \hat{S}_i \hat{H_u} \hat{H_d}$ will be important. On the other hand, for
$m_1^2 + m_2^2 < 0$ the $S_i$ will acquire large expectation values
along the flat direction at a scale intermediate between the electroweak and Planck
scales.  (The apparent runaway nature of the potential  in (\ref{flatd}) may be stabilized by loops
(i.e., the running of the $m_i^2$) or by higher-dimensional terms in 
the superpotential~\cite{intermediate}.) In the intermediate scale case, small
 Dirac neutrino (or other fermion) masses\footnote{Variations can yield comparable small
 Dirac and Majorana mass terms, as are needed to have mixing between ordinary
 and sterile neutrinos~\cite{sterile}.} may be generated by higher-dimensional
 terms such as
 \beq
W \sim \hat{H}_2 \hat{L}_L \hat{\nu}^c_L
            \left( {\hat{S}\over {M_{pl}}} \right)^{P_{D}}  \eeq
in the superpotential, where $P_D > 0$.
Possible cosmological implications of light Dirac neutrinos in \upr \ models are considered
in~\cite{BLL}.

\subsection{Family Structure and the Fermion Spectrum}
The family structure and fermion spectrum is perhaps the most
mysterious feature of the standard model. Attempts to shed light on
these issues in the framework of four-dimensional theories
often invoke family symmetries and their breakings to restrict the
form of the fermion mass matrices, often in conjunction with
grand unification (generally extended to allow higher dimensional Higgs representations).
Bottom-up models involving extra dimensions greatly extend the possibilities
for selection rules and family hierarchies.

String constructions allow another possibility, i.e., that string selection rules and
differences in  the embeddings for the three families (especially the third),
rather than symmetries in the four-dimensional effective theory, may
be the ultimate origin of the family structure. For example,
selection rules in heterotic constructions may lead to the absence
of couplings that are allowed by the symmetries of the effective
four-dimensional field theory. It is possible that such stringy effects
may ultimately explain the hierarchy of fermion masses and the pattern of their
mixings (and possibly imply new flavor changing effects, e.g., from new \zpr \ interactions).
Unfortunately, no construction that is very realistic in this respect has yet emerged.

Some issues that string model builders should keep in mind include:
the sources and magnitudes of CP-violating phases; possible
resolutions of the strong CP problem, such as Peccei-Quinn symmetries; 
and possibilities in the superpotential or elsewhere
for Majorana neutrino masses (in particular, whether terms leading to
Majorana masses  can be diagonal in neutrino flavor or only mix different
families).

The effective low energy theory in string constructions may be of the WYSINWYG
(what you see is {\em not} what you get) variety. For
 example, the number of families may be reduced from the
 apparent number in the compactification scheme by
 vacuum restabilization~\cite{clew}.  On the other hand,
 more complicated symmetry breaking patterns may increase the number
 of families\footnote{For example, a construction involving $SU(6)  \supset SU(2)^3$
 with a single 6-plet familiy could break to the diagonal $SU(2)$ with {\em three} 
 $SU(2)$-doublet families.}, or some  low energy particles may
 be composite, as occurs in an intersecting brane construction~\cite{CLW}.

\subsection{Asymptotic Freedom in the Quasi-Hidden Sector}
\label{qhs}
An important issue is whether the quasi-hidden sector is asymptotically free,
as is the case for the intersecting brane constructions in~\cite{CSU} (but not for the
heterotic constructions in~\cite{CCEELW}). An asymptotically free quasi-hidden
sector is a candidate for supersymmetry breaking via gaugino condensation~\cite{gaugino}
at the scale at which the coupling becomes strong, as well as
for dilaton/moduli stabilization.
For example, the dilaton and shape moduli are stabilized in the interesection
brane construction~\cite{CLW}, leading to the interesting features of
gaugino masses that are non-universal and which have large CP-violating phases.
However, the model is not realistic in that there is a large negative cosmological constant
and a large supersymmetry breaking scale around $10^{13}$ GeV.
Further study of such constructions may be useful in motivating new 
patterns of soft supersymmetry breaking, even if the constructions are not fully realistic. 

The strongly coupled groups may also lead to charge confinement and compositeness.
For example, in the construction in~\cite{CSU} there are states with non-standard
electric charge, such as $1/6$ or $\pm 1/2$. These all carry nontrivial charges under
the three quasi-hidden sector factors, and are presumably confined below
the strong coupling scales. Standard anomaly matching conditions then
require the existence of composite bound states, leading to the interesting feature
that the low energy spectrum contains an exotic fourth family (all elements are
\sto \ singlets) in which the left-chiral components are composite and the
right-chiral components elementary~\cite{CSL1}.

\section{Conclusions}
The Standard Model (extended to include neutrino masses) is extremely successful.
 Most aspects have been tested; in particular, precision electroweak data suggest
that any underlying new physics should be of the
decoupling type (i.e., the effects become smaller for larger mass scales),
such as supersymmetry and unification. Nevertheless, the Standard Model is
 clearly incomplete: it involves too many free parameters, arbitrary features, and
 fine tunings.
Superstring/M theory is an extremely promising theoretical direction, but testing it 
and picking from the large variety of vacua is extraordinarily challenging. 
We certainly  need a vigorous program of bottom-up experimental and theoretical probes
 to test the SM (or MSSM) and search for alternatives or extensions.
 However, it is also important to carry out a major
 top-down program to attempt to connect M theory to experiment and suggest
new TeV-scale physics that might not be motivated by bottom-up constructions.
There may be much beyond the  MSSM at the TeV scale.

\section*{Acknowledgments}
I am grateful to a number of individuals, including M. Cveti\v c, L. Everett, J. Wang,
 J. R. Espinosa, M. Pl\" umacher, J. Erler, T. Li, and G. Shiu, for enjoyable collaborations.
 This work was supported in part by a Department of Energy grant DOE-EY-76-02-3071.

%\begin{thebibliography}{99}
%\bibitem{ja}C Jarlskog in {\em CP Violation}, ed. C Jarlskog
%(World Scientific, Singapore, 1988).
%
%\bibitem{ma}L. Maiani, \Journal{\PLB}{62}{183}{1976}.
%
%\bibitem{bu}J.D. Bjorken and I. Dunietz, \Journal{\PRD}{36}{2109}{1987}.
%
%\bibitem{bd}C.D. Buchanan {\it et al}, \Journal{\PRD}{45}{4088}{1992}.
%
%\end{thebibliography}

\end{document}